\documentclass[preprint,floatfix,superscriptaddress,nofootinbib]{revtex4-2}

\usepackage{algorithm}
\usepackage{algpseudocode}
\usepackage{amsmath}
\usepackage{amsthm}
\usepackage{amssymb}
\usepackage{amscd}
\usepackage{bm}
\usepackage{booktabs}
\usepackage{latexsym}
\usepackage{mathtools}
\usepackage{comment}

\usepackage{graphicx}
\usepackage{epsfig}
\usepackage{epstopdf}
\usepackage{svg}
\usepackage[caption=false]{subfig}
\usepackage{placeins}
\usepackage{tabularx}

\usepackage{url}
\usepackage{hyperref}

\providecommand{\about}[0]{\raise.17ex\hbox{$\scriptstyle\sim$}}

\providecommand{\diff}{d}
\providecommand{\deriv}[3][ ]{\frac{\diff^{#1}#2}{\diff #3^{#1}}}

\providecommand{\eqref}[1]{(\ref{#1})}

\providecommand{\vanish}[1]{}

\graphicspath{{./figures/}}

\begin{document}


\title{Energy storage under high-rate compression of single crystal tantalum}

\author{J. C. Stimac}
\email{jstimac@ucdavis.edu}
\affiliation{Department of Chemical Engineering, University of California, Davis, CA, 95616, USA.}

\author{N. Bertin}
\email{bertin1@llnl.gov}
\affiliation{Lawrence Livermore National Laboratory, Livermore, CA, 94550, USA.}

\author{J. K. Mason}
\email{jkmason@ucdavis.edu}
\affiliation{Department of Materials Science and Engineering, University of California, Davis, CA, 95616, USA.}

\author{V. V. Bulatov}
\email{bulatov1@llnl.gov}
\affiliation{Lawrence Livermore National Laboratory, Livermore, CA, 94550, USA.}

\begin{abstract}
When a material is plastically deformed the majority of mechanical work is dissipated as heat, and the fraction of plastic work converted into heat is known as the Taylor-Quinney coefficient (TQC).
Large-scale molecular dynamics simulations were performed of high strain rate compression of single-crystal tantalum, and the resulting integral and differential TQC values are reported up to true strains of $1.0$.
A phenomenological model is proposed for the energy stored in the material as a function of time with an asymptotic limit for this energy defined by the deformation conditions.
The model reasonably describes the convergence of TQC values to $1.0$ with increasing plastic strain, but does not directly address the physical nature of thermo-mechanical conversion.
This is instead developed in a second more detailed model that accurately accounts for energy storage in two distinct contributions, one being the growing dislocation network and the other the point defect debris left behind by the moving dislocations.
The contribution of the point defect debris is found to lag behind that of the dislocation network but to be substantial under the high-rate straining conditions considered here.

\end{abstract}


\maketitle

\section{Introduction}
\label{sec:introduction}

The Taylor-Quinney coefficient (TQC), or the coefficient of thermo-mechanical conversion, is named for two scientists who observed that only a fraction of the mechanical work performed on a plastically deformed material is converted into heat and concluded that the remainder must be stored in the material \cite{1934Taylor}.
The TQC is specifically defined as the fraction of mechanical work converted into heat, with the difference between one and the TQC indicating the extent of energy storage.
The original experiments by Taylor and Quinney imposed significant torsion and compression on copper rods and found that this ratio began around $0.85$ or $0.9$ before converging to $1.0$ with increasing strain.
As a consequence, the TQC is frequently approximated as a constant around $0.9$.
Although its nature was unknown at the time, Taylor and Quinney convincingly related the stored energy to an increase in the flow stress (plastic strength), a phenomenon that would eventually become known as strain hardening.
This suggested that perhaps the stored energy itself constituted a material state variable that evolved during plastic deformation and defined the material strength.
Less than a year after publishing the original paper, Taylor further developed this connection by proposing (simultaneously with Orowan and Polanyi) that crystal plasticity results from the motion of distinct line-shaped defects called dislocations \cite{1934Taylor_2}.
Following that ground-breaking discovery nearly $90$ years ago, it has subsequently been established that the stored energy reflects changes in the defect microstructure, e.g., the distributions of dislocations, point defects, grain boundaries, stacking faults, deformation twins, etc.
For most deformation conditions of interest though, both strain hardening and energy storage are mainly consequences of the multiplication and storage of dislocations.

From a theoretical standpoint, it has been proposed that the TQC is related to the configurational entropy.
An early review made note of estimated contributions to this entropy change from dislocations and point defects to estimate the free energy change during a cold-work process \cite{1973Bever}.
Later, the TQC became a quantity of interest in thermodynamic theories of plasticity \cite{berdichevsky2006thermodynamics}, including the thermodynamic dislocation theory (TDT) \cite{langer2010thermodynamic, langer2019statistical} which introduced an effective temperature to quantify the configurational disorder of non-equilibrium dislocation systems.
This later framework was subsequently used to derive thermodynamic constraints on the TQC \cite{2020Lieou}.
Recent work by Zubelewics also proposed a novel interpretation of the TQC as a measure of the ability of the material to raise the configurational entropy enough to untangle the dislocation network and enable dislocation motion \cite{2019Zubelewicz}.

Experimentally, direct measurement of the TQC requires delicate accounting of the heat released by the material during the prescribed straining conditions.
The measurement accuracy is hampered at low strain rates by the typically small amounts of heat released and the difficulty of thermally isolating the system from its environment.
High strain-rate deformation can release substantially more heat and suffers from fewer thermal losses to the environment, but can make thermal measurements themselves more difficult.
For example, measurements of the dissipated thermal energy by means of infrared detection have been found to be rather sensitive to the calibration procedure, with updates to this procedure able to change the apparent TQC by as much as a factor of two \cite{2000Macdougall}.
During material tests performed at most extreme deformation rates, e.g., at the National Ignition Facility \cite{fratanduono2021,kraus2022}, uncertainty in specimen temperature can reach hundreds of Kelvin making extraction of the material’s intrinsic thermo-mechanical properties quite difficult.
Despite (or perhaps due to) our continually expanding and improving experimental capabilities, the range of TQC values appearing in the literature is not narrowing, and if anything has expanded to the entire interval between $0.1$ and $1.0$.
Reference \cite{2017Rittel} provides a review of the recent literature on this increasingly controversial subject.

Accurate knowledge of the TQC is desirable since it allows the prediction of the temperature change of a material that is plastically deforming in adiabatic conditions, e.g., at high strain rates.
Given that a material's mechanical properties can be temperature dependent, the TQC can appear as a critical parameter in continuum-level models of plastic deformation \cite{1993Mason,2021Kositski}.
The value of the TQC can be relevant to other simulations of material behavior as well, with the stored energy of cold work known to be implicated in the onset of recrystallization and the Baushinger effect \cite{benzerga2005stored}.

Since it remains effectively impossible to partition mechanical work between released heat and stored energy in experimental tests performed at the most extreme deformation conditions, computer simulations offer an alternative means to gain much needed insights into energy storage.
Unlike high strain rate experiments, numerical simulations can provide complete information about the simulated material response and allow mechanical work, released heat and stored energy to all be computed and monitored incrementally at every time step.
Reference \cite{benzerga2005stored} was perhaps the first simulation to explore energy storage due to dislocation multiplication during plastic deformation, and used a discrete dislocation dynamics (DDD) method.
However, given that it was a highly abridged two-dimensional version of DDD with several critical parameter values adjusted in an ad hoc fashion to compensate for the missing third dimension, the values of the TQC reported there do not represent any real material or deformation condition.
More recently, molecular dynamics (MD) simulations of dislocation motion in single crystals and of plasticity in nanocrystals \cite{2021Kositski} were used to calculate the TQC of aluminum, copper, iron, and tantalum under high strain-rate shearing in isothermal-isobaric conditions.
Based on these simulations, the authors concluded that only extreme dislocation multiplication up to densities of $10^{16} \ \mathrm{m^{-2}}$ or extreme grain refinement down to diameters of $10 \ \mathrm{nm}$ could account for a TQC lower than $1.0$ at the deformation rates accessible to MD simulations (${\about} 10^8 \ \mathrm{s^{-1}}$).
The authors then argued that grain refinement is the more likely of the two mechanisms to account for appreciable energy storage in practice.
That said, their simulations of nanograined samples had grain sizes that were orders of magnitude smaller than what is usually observed in experimental specimens, artificially inflating the apparent contribution of grain refinement to energy storage.  Furthermore, their simulations of dislocated single crystals initially contained only a single edge dislocation dipole, effectively precluding dislocation multiplication as a means of energy storage.

In this paper, large-scale MD simulations of high-rate deformation of single crystal tantalum in isothermal-isochoric conditions are performed and analyzed up to a true strain of $1.0$ to observe and quantify the role of dislocation multiplication as a mechanism of energy storage.
Tantalum (Ta) has long been recognized for its toughness and resistance to corrosion \cite{1989Kock}, resulting in its use in a variety of high strain-rate applications.
For this reason, Ta is used as a model material to study body-centered cubic (BCC) metals under extreme strains and strain rates.
The MD simulations and the methods used to monitor the thermo-mechanical response of simulated single crystals are described in Sec.\ \ref{sec:methods}.
Our simulation results on the TQC and energy storage in Ta for the specified deformation protocols are given in Sec.\ \ref{sec:results}.
The relative utility of the differential and integral forms of the TQC for understanding the materials response, a phenomenological model for both the differential and integral TQC as functions of time, and a model that describes the energy stored in the defect microstructure as a function of only the dislocation density and point defect concentration are all discussed in Sec.\ \ref{sec:discussion}.
Conclusions on the basis of our observations are offered in Sec.\ \ref{sec:conclusion}.

\section{Methods}
\label{sec:methods}

All simulations were carried out using the LAMMPS software \cite{LAMMPS} with the embedded atom method (EAM) interatomic potential for Ta of Li et al.\ \cite{2003Li} and a $10 \ \mathrm{fs}$ time step.
A single-crystal rectangular prism of Ta containing $\about 33$ million atoms was generated with a 2:1:1 aspect ratio along the $[100]$, $[010]$ and $[001]$ crystallographic directions.
\begin{figure}
	\centering
	\includegraphics[width=0.80\textwidth]{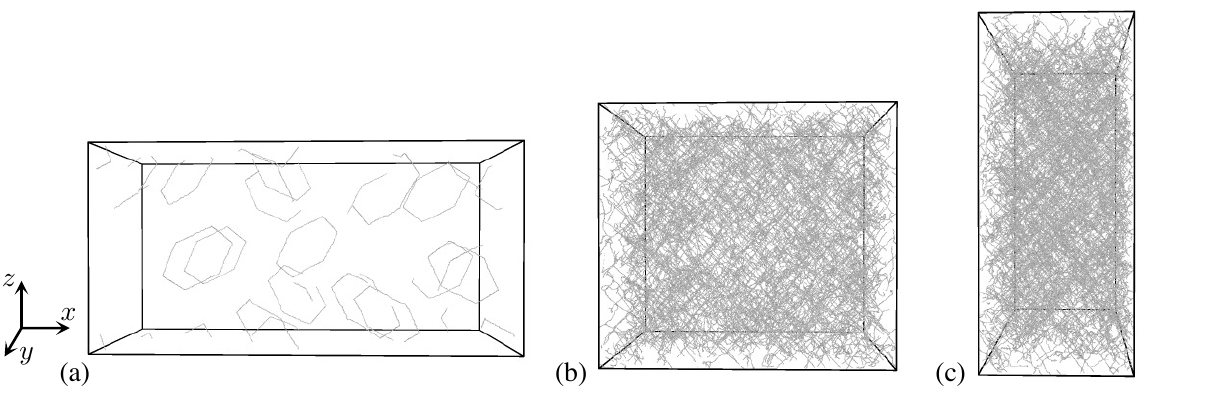}
	\caption{Simulation volume and  dislocation network at true plastic strains of (a) $0.0$, (b) $0.4$, and (c) $1.0$.
	The compression axis $x$ points to the right in all three frames.} 
	\label{fig:sim_box_geom}
\end{figure}
Periodic boundary conditions were enforced in all directions, and the crystal was seeded with an equal number of hexagonal dislocation loops for each of the four $\frac{1}{2}\langle111\rangle$ Burgers vectors of the body-centered cubic lattice as is visible in Fig.\ \ref{fig:sim_box_geom}a.
The crystal was then subject to uniaxial compression along the $x$-dimension (the longest dimension aligned with the $[100]$ direction), and the displacements along the $y$- and $z$-dimensions enforced that the deformed crystal remain a tetragonal prism of constant volume.
The temperature was maintained at $300\ \mathrm{K}$ by means of a Langevin thermostat.
All simulation settings were identical or similar to those reported in Ref.\ \cite{zepeda2017probing}, the only difference being the use of isothermal-isochoric rather than isothermal-isobaric conditions.
This minor change was made largely for computational convenience since the crystal volume was fixed to maintain pressures below $200\ \mathrm{MPa}$ during plastic flow (crystal plasticity is naturally isochoric), and several smaller simulations comparing isochoric and isobaric conditions exhibited virtually identical responses.

The crystal was subjected to two different dislocation seed configurations and deformation protocols to investigate the extent to which the TQC depends on the initial dislocation density and loading conditions.
The first simulation (A) initialized the crystal with $12$ hexagonal dislocation loops that were $40$ lattice constants in diameter and subjected the crystal to a constant compressive strain rate along the $x$-axis of $2 \times 10^{8}\ \mathrm{s^{-1}}$ throughout the entire simulation.
As discussed below, this caused a stress overshoot to develop as a consequence of the initial few dislocations requiring time to multiply to a level sufficient to relieve the stress by means of their motion.
The second simulation (B) initialized the crystal with $16$ hexagonal dislocation loops that were $80$ lattice constants in diameter, twice that of the first simulation.
The compressive strain rate along the $x$-axis was linearly ramped from $0$ to $4 \times 10^{8}\ \mathrm{s^{-1}}$ over an interval of $2\ \mathrm{ns}$ to a total strain of $0.4$, and afterwards held constant at $4 \times 10^{8}\ \mathrm{s^{-1}}$.
As a result of ramping the strain rate and initializing a greater number of larger-sized dislocation loops, simulation B did not exhibit a stress overshoot.

\begin{figure}
	\centering
	\includegraphics[width=0.45\textwidth]{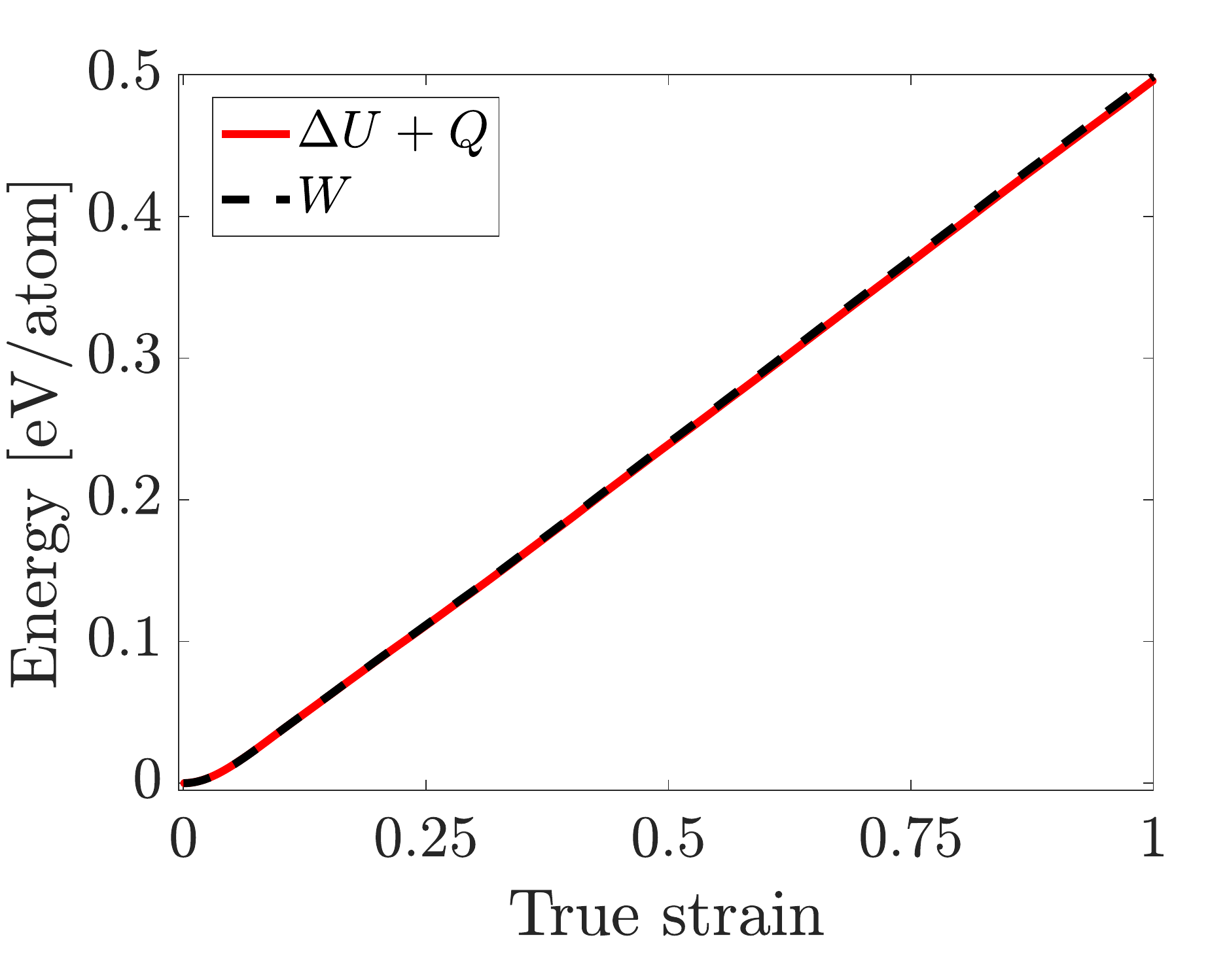}
	\caption{Verification that the ramped simulation obeys the first law of thermodynamics. The total mechanical work $W$ done on the crystal is equal to $\Delta U + Q$, or the sum of the change in the internal energy $\Delta U$ and the total heat $Q$ exchanged with the thermostat.}
	\label{fig:conserved_energy}
\end{figure}

The amount of external mechanical work $dW$ performed on the crystal in a given time step was computed as
\begin{equation}
    dW = \sum_i \sigma_{ii} A_{i} dL_{i}
    \label{eq:differential_work}
\end{equation}
or as the sum of the products of the diagonal components $\sigma_{ii}$ of the internal stress tensor, the areas $A_{i}$ of the corresponding faces, and the incremental displacements $dL_{i}$ of the faces in their normal directions (the shear components were all zero since the simulation volume remained strictly tetragonal).
The mechanical work, the internal energy $U$ of the simulation cell, and the incremental thermal energy $dQ$ exchanged with the Langevin thermostat were continuously recorded during the simulation.
Figure \ref{fig:conserved_energy} shows that, to a high degree of accuracy and over the entire simulation, the total mechanical work $W$ performed on the crystal (the integral of Eq.\ \ref{eq:differential_work}) is equal to the sum of the change in the internal energy $\Delta U$ and the total heat $Q$ exchanged with the thermostat.
Our partitioning of mechanical work into dissipated heat and stored energy should be accurate to the extent that this energy balance holds.

\begin{figure}
	\centering
	\includegraphics[width=0.6\textwidth]{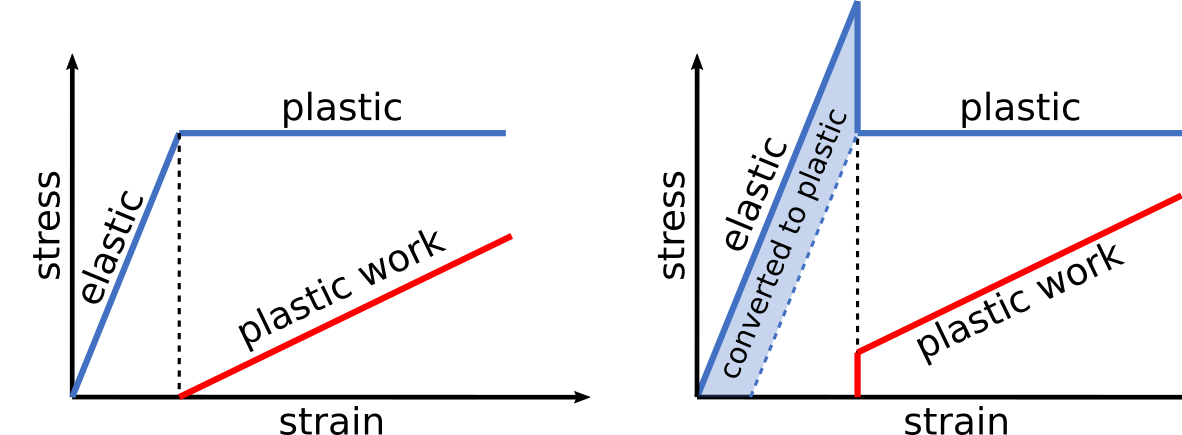}
	\caption{Stress-strain schematics that indicate the relationship of yield point phenomenon and the transformation of elastic work to plastic work at low strains.}
	\label{fig:stress_overshoot}
\end{figure}

In their original paper, Taylor and Quinney computed the work-to-heat conversion ratio by dividing the heat released by the total mechanical work, the latter computed as the area under the true stress--true strain curve \cite{1934Taylor}.
It has subsequently become more common to use the plastic work instead of the total work, the rationale being that the elastic component is reversible and therefore never converted into heat.
This is schematically illustrated on the left side of Fig.\ \ref{fig:stress_overshoot} where plastic work only begins to accrue after the onset of yielding.
This distinction has no bearing on the results and observations in the original paper since, at the low deformation rates employed there, the elastic work constituted a negligible fraction of the total work.
However, at the high compressive strain rates employed in our MD simulations the crystal experienced much higher stresses on the order of several GPa (and even higher during the stress overshoot in simulation A).
Thus, a substantial fraction of the total work is stored in the form of elastic strain energy.
Since our concern is with the part of the total work that is irreversibly stored in the defect microstructure, here we follow the prevailing practice and use only the plastic work to calculate the TQC.

The plastic work is found by subtracting the elastic energy (the energy that could be recovered by unloading the crystal) from the total work.
The equation for the elastic energy for cubic crystals in the $\langle100\rangle$ frame reduces to
\begin{align}
    E_\mathrm{el} &= \frac{V}{2} \sum_{ijkl} S_{ijkl} \sigma_{ij} \sigma_{kl} \nonumber \\
    &= \frac{V}{2} [S_{11}(\sigma_{11}^2 + \sigma_{22}^2 + \sigma_{33}^2) + 2 S_{12}(\sigma_{11}\sigma_{22} + \sigma_{11}\sigma_{33} + \sigma_{22}\sigma_{33}) + S_{44}(\sigma_{12}^2 + \sigma_{13}^2 + \sigma_{23}^2)]
    \label{eq:elastic_energy}
\end{align}
where $V$ is the crystal volume, $\sigma_{ij}$ are the internal stress tensor components, and $S_{ijkl}$ and $S_{ij}$ are the elastic compliance components in tensor and Voigt notations.
The elastic compliance components were found from the elastic stiffness components $C_{ij}$ by the standard equations
\begin{align}
    S_{11} &= \frac{C_{11} + C_{12}}{(C_{11} - C_{12})(C_{11} + 2 C_{12})} \nonumber \\
    S_{12} &= \frac{-C_{12}}{(C_{11} - C_{12})(C_{11} + 2 C_{12})} \nonumber \\
    S_{44} &= \frac{1}{C_{44}}
\end{align}
and the elastic stiffness components were measured in a separate calculation using the same interatomic potential, temperature and pressure as in simulations A and B.

\begin{figure}
	\centering
	\includegraphics[width=0.45\textwidth]{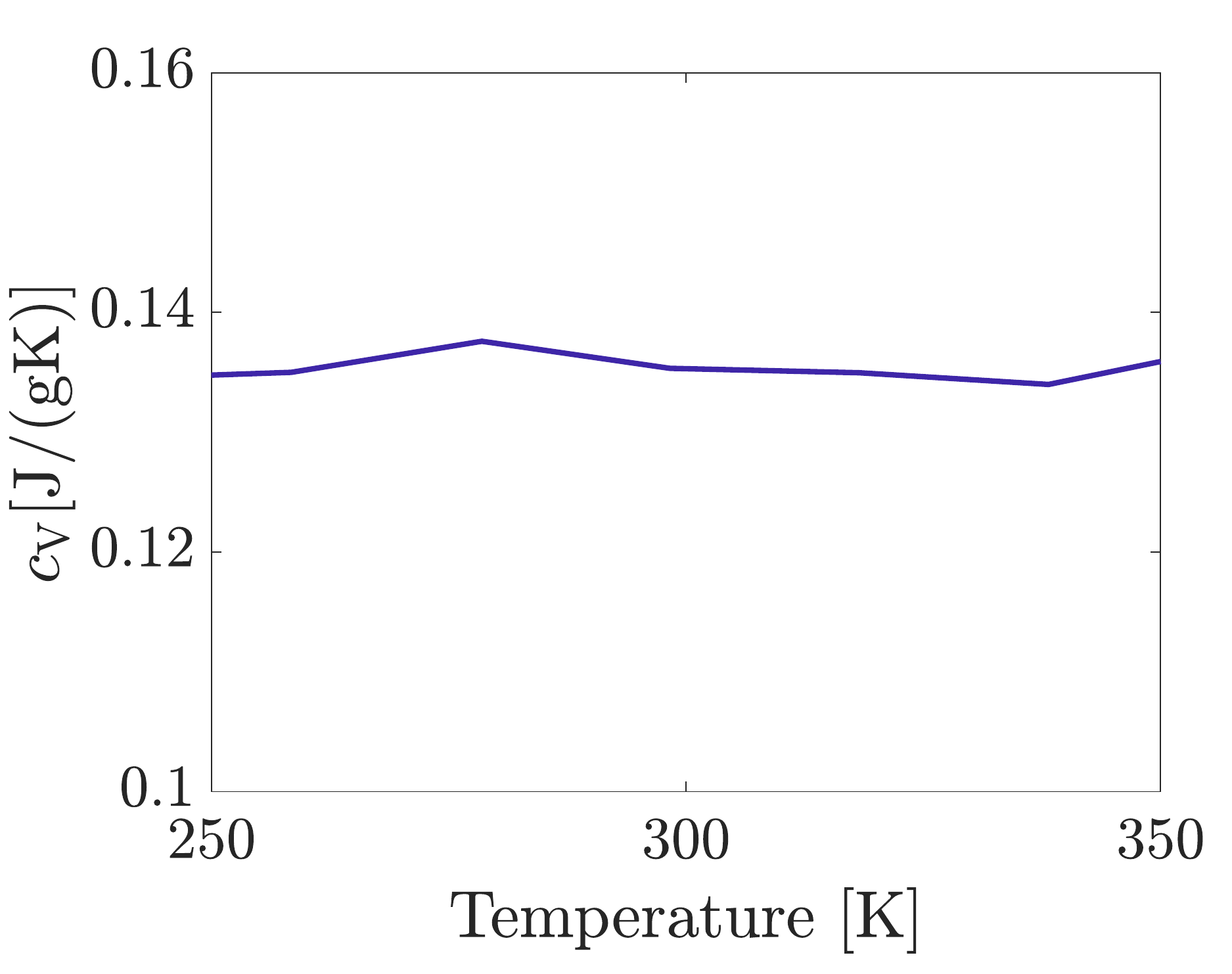}
	\caption{Constant-volume specific heat capacity of Ta over the relevant temperature range.
	The blue trend interpolates observations measured over $1\ \mathrm{ps}$ intervals.}
	\label{fig:Cv}
\end{figure}

The Langevin thermostat employed in our simulations did not strictly maintain the temperature at $300\ \mathrm{K}$, and instead allowed the temperature to increase by a few Kelvin.
Therefore, a precise calculation of the fraction of mechanical work expended in forming defect structures requires knowledge of the heat capacity of Ta for the selected interatomic potential \cite{2003Li}.
The specific heat capacity at constant volume was measured by maintaining the simulation cell at constant volume, slowly incrementing the target temperature over the interval $250\ \mathrm{K} \leq T \leq 350\ \mathrm{K}$, and continuously observing the heat exchanged with the thermostat. 
The results are reported in Fig.\ \ref{fig:Cv} where the specific heat capacity at constant volume apprears to stay constant at $0.135\ \mathrm{J/(g K)}$ in the relevant interval, coincidentally agreeing with the $c_{p} = 0.14\ \mathrm{J/(g K)}$ reported in Ref.\ \cite{2021Kositski} for a different interatomic potential for Ta.

\section{Results}
\label{sec:results}

\begin{figure}
	\centering
	\includegraphics[width=0.9\textwidth]{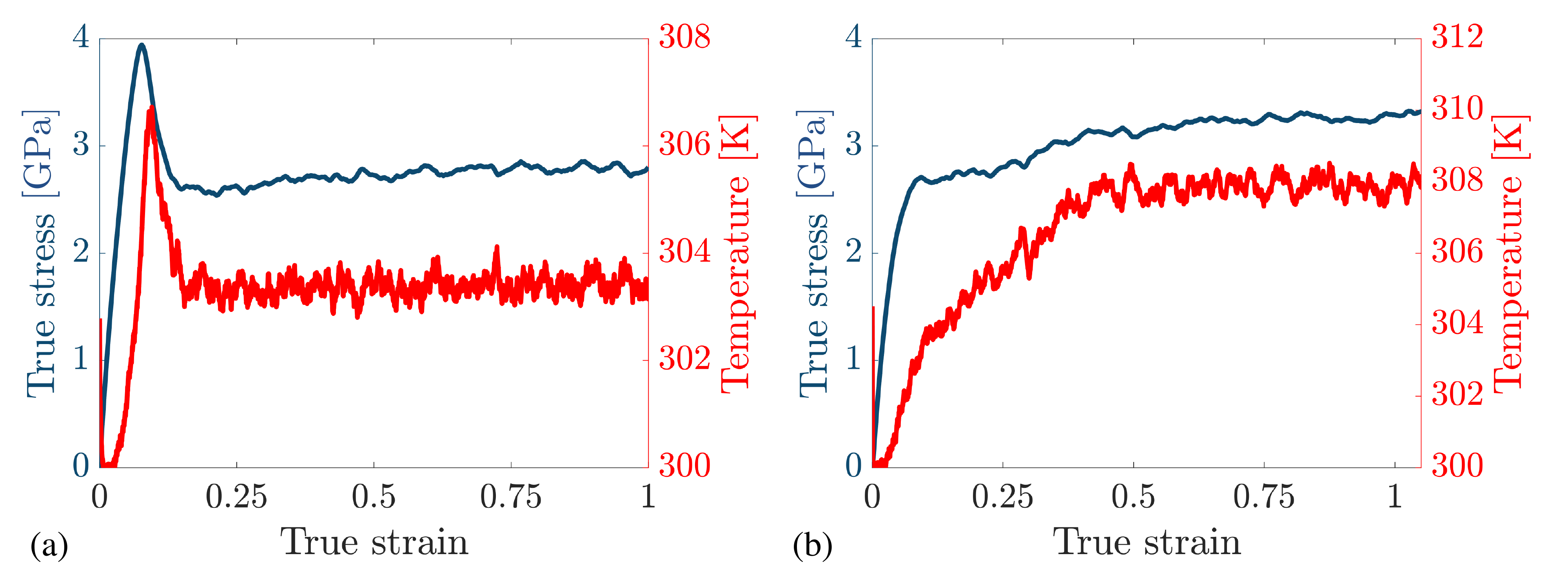}
	\caption{True stress and absolute temperature as a function of true strain for (a) simulation A and (b) simulation B.
	While the temperature is not constant, the Langevin thermostat keeps the temperature to within a few Kelvin of the target temperature, with the difference being roughly proportional to the instantaneous rate of plastic work.}
	\label{fig:stress_temp_strain}
\end{figure}

The stress in simulation A (with a constant strain rate) is shown in Fig.\ \ref{fig:stress_temp_strain}a and exhibits an overshoot to an upper yield point followed by a substantial drop to a lower yield point, with the flow stress afterwards remaining nearly constant for the rest of the simulation.
Particularly at high strain rates, the initial transient in such stress-strain curves is generally attributed to delayed dislocation multiplication in response to a rising stress.
The initially low dislocation density limits the amount of deformation that can be accommodated via dislocation motion, resulting in the majority of the mechanical work being converted into elastic energy.
On reaching the upper yield point, dislocations have multiplied sufficiently for their motion to produce plastic strain at a rate that equals and even exceeds the externally-imposed strain rate;
this overproduction of plastic strain causes the subsequent stress drop.
As the stress drops, the elastic energy accumulated during the initial stress rise is rapidly converted into plastic work (shaded area in the schematic on the right of Fig.\ \ref{fig:stress_overshoot}), most of which is immediately released as heat.
This temporarily increased the temperature to $307\ \mathrm{K}$ despite the use of a Langevin thermostat intended to maintain the temperature constant at $300\ \mathrm{K}$.
The thermostat relaxation time was set to $10\ \mathrm{ps}$ as a compromise between the desires for a constant simulation temperature and to avoid excessively perturbing the dynamics from the random atomic kicks and friction forces that the thermostat employs.
The majority of the heat released during the simulation is still exchanged with the thermostat, with the minority responsible for the observed temperature variations being accounted for by adding the term $C_{V} dT(t)$ to the incremental heat exchanged with the reservoir $dQ(t)$.
The specific heat capacity $0.135\ \mathrm{J/(g K)}$ (Fig.\ \ref{fig:Cv}) and mass density $16.55\ \mathrm{g/cm^{-3}}$ were specific to the Ta model used in this study.
This correction affects our computed TQC only during the initial yield transient and becomes insignificant past the lower yield point where the temperature settles at about $304\ \mathrm{K}$ and stays very nearly constant through the rest of the simulation.

The intent with simulation B was to suppress the initial upper-lower yield transient that caused the rapid conversion of elastic energy into plastic work after the upper yield point in simulation A, thus possibly affecting the value of the TQC.
This was achieved (Figure \ref{fig:stress_temp_strain}b) by seeding a greater initial dislocation density and linearly increasing (ramping) the strain rate.   
The temperature in simulation B similarly rises monotonically with strain, though to a higher value than in simulation A as is consistent with the higher steady-state strain rate.

The differential TQC is defined as the ratio of the incremental heat released to incremental plastic work performed during the time step starting at time $t$, or \cite{1993Mason,rittel1999conversion}
\begin{equation}
    \beta_\mathrm{diff}(t) = \frac{dQ(t) + C_{V} dT(t)}{dW_{p}(t)}
    \label{eq:beta_diff}
\end{equation}
where $dW_{p}(t) = dW(t) - dE_\mathrm{el}(t)$ is the incremental plastic work defined by Eqs.\ \ref{eq:differential_work} and \ref{eq:elastic_energy} and the remaining quantities are taken directly from the standard output of the MD simulation.
The integral TQC is instead defined as the ratio of total heat released by time $t$ and the total plastic work done on the crystal by the same time $t$:
\begin{equation}
    \beta_\mathrm{int}(t) = \frac{Q(t) + C_{V} \Delta T(t)}{W_{p}(t)}.
    \label{eq:beta_int}
\end{equation}
While $\beta_\mathrm{diff}$ is more informative as a measure of the instantaneous partitioning of mechanical work along the deformation path, $\beta_\mathrm{int}$ is more commonly reported of the two in the literature, perhaps for the reason that it is easier to measure experimentally.
As was explained in \cite{rittel1999conversion} where the two parameters were first referred to as ``differential'' and ``integral'', the integral TQC is not equal to the integral of the differential TQC.
Instead, the following equality holds:
\begin{equation}
    \beta_\mathrm{int}(t) = \int_{0}^{t} \beta_\mathrm{diff}(t') \frac{dW_{p}(t')}{W_{p}(t)}   
    \label{eq:beta_conversion}
\end{equation}
in which the contribution of $\beta_\mathrm{diff}$ computed at time $t'$ to $\beta_\mathrm{int}$ is weighted by the fractional plastic work performed during $[t', t' + dt']$ relative to that during $[0, t]$.

\begin{figure}
	\centering
    \includegraphics[width=0.9\textwidth]{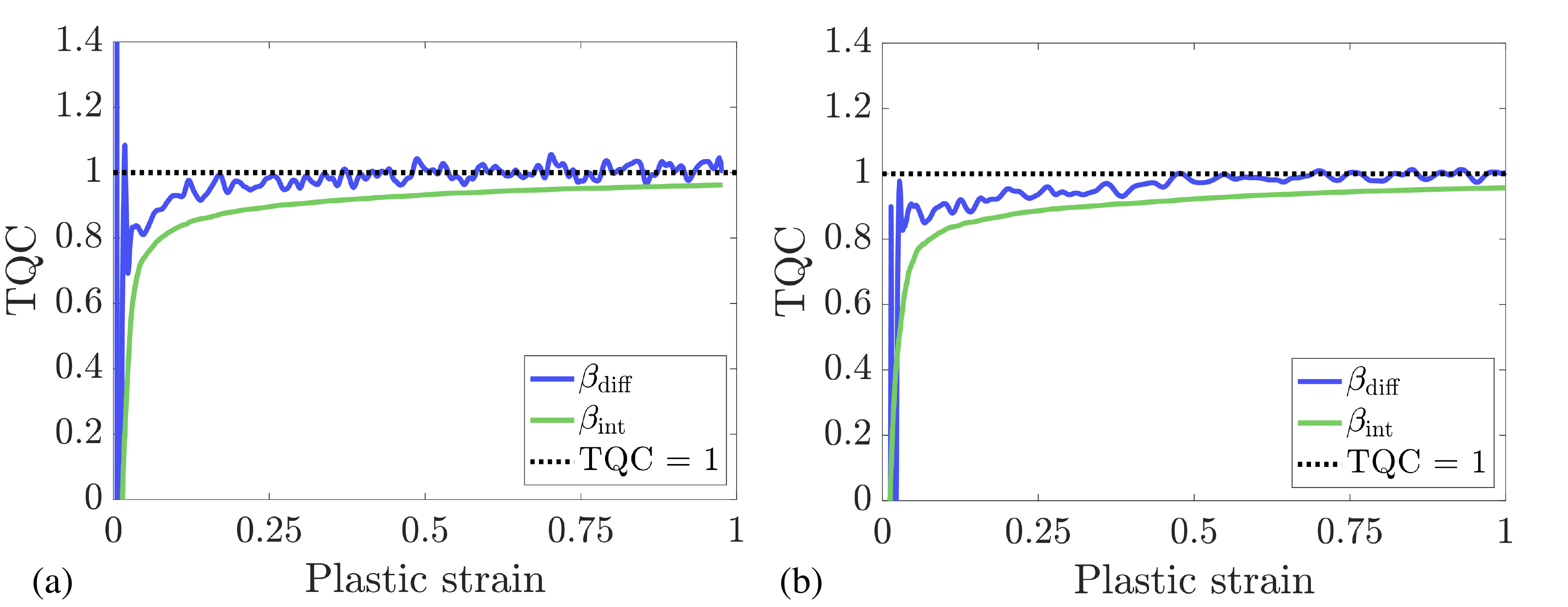}
	\caption{Differential (blue) and integral (green) TQC as calculated by Eqs.\ \ref{eq:beta_diff} and \ref{eq:beta_int} for (a) constant (simulation A) and (b) ramped (simulation B) strain-rate compression.}
	\label{fig:TQC}
\end{figure}

Figure \ref{fig:TQC} shows the computed differential and integral TQC values as blue and green lines for simulations A and B.
The differential TQC is susceptible to high-frequency fluctuations attributed to the variability of the small values of the denominator in Eq.\ \ref{eq:beta_diff} computed over short time intervals, especially during the initial stress rise where only a small fraction of the mechanical work is converted into plastic work;
such variability can even cause $\beta_\mathrm{diff}$ to be ostensibly higher than one over a few time intervals.
Indeed, fluctuations of this kind are all too familiar in MD simulations with finite numbers of atoms.
For the systems of the size considered here, a smoothing operation is applied to the differential TQC that entails a moving average in a $20 \ \mathrm{ps}$ window iteratively applied for $10$ passes.
The rapid increase of the integral and differential TQC during and after yielding indicates that initially a large fraction of plastic work is expended in generating new crystal defects (storage), and that this fraction gradually decreases with increasing plastic strain as the system approaches a state of stationary flow.

\section{Discussion}
\label{sec:discussion}

Simulations A and B are generally consistent with each other and with other similar large-scale simulations reported elsewhere \cite{zepeda2017probing, zepeda2021atomistic,bertin2022sweep}.
As has been observed in all simulations performed at comparable strain rates, the flow stress reaches a plateau at true strains in the range of $0.3$ to $0.5$ (Fig.\ \ref{fig:stress_temp_strain}).
The flow stress in simulation B has nearly plateaued when the end of ramping is reached at a true strain of $0.4$ and remains nearly constant from a true strain of $\about 0.5$ to the end of the simulation.
The plateau stress is higher than in simulation A because the final strain rate of $4 \times 10^{8} \ \mathrm{s^{-1}}$ in simulation B is twice as high as that in simulation A.

\subsection{Comparison between TQC measures}
\label{subsec:comparison_TQC}

Simulations A and B confirm earlier observations that, when subjected to straining under conditions that remain constant for a sufficiently long time, a plastic crystal can attain a state of stationary flow \cite{zepeda2017probing}.
In such a stationary state the dislocation microstructure itself becomes stationary but only in a statistical sense;
although mutually balanced, dislocation multiplication and annihilation continue unabated resulting in the incessant rebuilding and dissolution of a dynamically evolving dislocation network.
Since the total defect density eventually reaches a stationary state, $\beta_\mathrm{diff}$ must be $1.0$ in stationary flow---to the degree the defect density is unvarying---since all of the plastic work must therefore be dissipated as heat.
The heat is generated primarily by dislocation motion converting the work of the Peach-Kohler force into the energy of lattice vibrations (phonons).
Further plastic work is converted into phonons as a result of energy-reducing dislocation reactions such as dislocation annihilation (and to a lesser extent the formation of dislocation junctions).
In stationary flow, the energy released as heat through dislocation annihilation is compensated by the additional defect energy accrued in the crystal by dislocation multiplication, leaving the net defect energy invariant.

This is fully consistent with data presented in Fig.\ \ref{fig:TQC} showing that, indeed, on reaching the true strain where the flow stress reaches its plateau, $\beta_\mathrm{diff}$ reaches and remains close to $1.0$ within the fluctuations expected in our simulations.
Although not obvious from Fig.\ \ref{fig:TQC}, the average of $\beta_\mathrm{diff}$ is actually slightly below $1.0$ as a consequence of the point defect density continuing to rise past the apparent start of stationary flow, discussed further in Sec.\ \ref{subsec:stored_energy}.
By contrast, $\beta_\mathrm{int}$ never reaches but only asymptotically approaches $1.0$ from below.
This is a direct consequence of the relationship in Eq.\ \ref{eq:beta_conversion} since, if $\beta_\mathrm{diff}$ is smaller than $1.0$ at any point and does not rise above $1.0$, $\beta_\mathrm{int}$ necessarily remains smaller than $1.0$ for all strains.
As occurred in simulations A and B, there is often appreciable energy storage at the early stages of straining that is integrated into $\beta_\mathrm{int}$, making this parameter rather insensitive to the subsequent straining trajectory.
For this reason it is tempting to advocate, after Ref.\ \cite{2021Kositski}, that $\beta_\mathrm{diff}$ is the more informative of the two parameters since it provides information about the instantaneous state of the system.
On the other hand, $\beta_\mathrm{diff}$ can be difficult or even impossible to extract during experiments, whereas $\beta_\mathrm{int}$ gives a simple measure of thermo-mechanical conversion under straining.
Regardless, their relationship in Eq.\ \ref{eq:beta_conversion} makes clear that the two quantities contain precisely the same information, simply in different forms.
Perhaps a more pertinent issue is that a value of the TQC (whether differential or integral) obtained in one experiment or simulation is generally not applicable to experiments or simulations performed under different conditions, depending as they do on the material's initial defect microstructure as well as on the entire thermo-mechanical trajectory followed during the test (discussed further in Sec.\ \ref{subsec:phenom} below).

Asymptotic convergence notwithstanding, our simulations indicate that $\beta_\mathrm{int}$ remains markedly lower than $1.0$ up to rather large plastic strains in situations where dislocation multiplication is the primary energy storage mechanism.
For instance, $\beta_\mathrm{int}$ is still below $0.9$ at a plastic strain of $0.5$ for both simulations in Fig.\ \ref{fig:TQC}.
This is at variance with Ref.\ \cite{2021Kositski} which found that $\beta_\mathrm{int}$ quickly converged to $1.0$ after plastic yield in MD simulations involving dislocations.
They concluded that energy storage due to dislocation multiplication could not explain experimental observations of $\beta_\mathrm{int}$ less than $1.0$, and that some other energy storage mechanism must be responsible.
It is significant to this discussion that the simulations in Ref.\ \cite{2021Kositski} used an initial condition consisting of a single edge dislocation dipole, effectively precluding dislocation multiplication which is the key energy storage mechanism in crystal plasticity.
By comparison, dislocation multiplication takes place naturally in our simulations in a statistically representative ensemble of interacting dislocations.
Indeed, our results suggest the opposite conclusion, that energy storage due to dislocation multiplication is substantial and should be accounted for when assessing thermo-mechanical conversion during high-rate straining experiments.
As discussed in more detail in Sec.\ \ref{subsec:stored_energy}, the total energy additionally stored in the defect microstructure in simulations A and B is 17 and 22 meV/atom, respectively, or roughly 5\% of the total plastic work performed on the crystals in the same simulations.
Assuming that the specific heat capacity remains unchanged at 0.135 $\mathrm{J/(g K)}$, this amounts to about $80$ K of additional temperature rise if all of the stored energy were to be released as heat.
By comparison, an estimate given in Ref.\ \cite{2021Kositski} for the amount of energy that could be stored by a grain refinement mechanism is only $2.5 \times 10^3\ \mathrm{J/m^3}$ or $0.28 \times 10^{-3}\ \mathrm{meV/atom}$, five orders of magnitude lower than what is observed in the two simulations reported here and hence entirely negligible on the energy scales relevant to high-rate plasticity.

\subsection{Phenomenological model}
\label{subsec:phenom}

As discussed in Sec.\ \ref{sec:introduction}, Taylor and Quinney conjectured in their original paper that the stored energy constituted a material state variable \cite{1934Taylor}.
They did this without knowledge of the defect microstructure, and while our understanding of such defects has developed extensively in the intervening years, there remains an appealing simplicity to modeling the evolution of the energy stored in the defect microstructure without concern for the specifics of the defect evolution processes (a subject explored in Sec.\ \ref{subsec:stored_energy}).

Let $E_\mathrm{stored} = U - U_\mathrm{ref} - E_\mathrm{el} - C_{V} \Delta T$ be defined as the energy stored in the defect microstructure, where $\Delta T$ is the deviation from the reference temperature ($300\ \mathrm{K}$) and $U_\mathrm{ref} = N_\mathrm{at} E_\mathrm{coh}$ where $N_\mathrm{at}$ is the number of atoms in the crystal and $E_\mathrm{coh}$ is the per-atom energy in a perfect crystal at the reference temperature.
Suppose that there is a limiting value of $E_\mathrm{stored}$ attained in a material that is continuously deformed for a sufficiently long time under fixed straining conditions.
Denoted $E_\mathrm{\infty}$, this asymptotic value is a material-specific function that presumably depends on temperature, pressure and straining rate.
If these control variables depend on time, then $E_\mathrm{\infty}$ becomes a function of time as well.
The difference $E_\mathrm{\infty}(t) - E_\mathrm{stored}(t)$ is then the remaining capacity of the material to store energy in the defect microstructure given the instantaneous straining conditions at time $t$.
Suppose further that during each time interval $dt$ the material stores in the defect microstructure a fraction of the incremental plastic work that is proportional to the remaining energy storage capacity, or
\begin{equation}
    d E_\mathrm{stored} / d t = \zeta (E_\infty(t) - E_\mathrm{stored}) d W_{p} / d t
\end{equation}
where $\zeta$ is the constant of proportionality (conjectured to be a material constant).
The general solution of this first-order linear ODE is 
\begin{equation}
    E_\mathrm{stored}(t) = \bigg\{ E_{0} + \int_0^t E_\infty(t') \zeta \deriv{W_{p}}{t'} \exp[\zeta W_{p}(t')] dt' \bigg\} \exp(-\zeta W_{p}(t))
\end{equation}
where $E_0 = E_\mathrm{stored}(0)$ is the energy stored in the defect microstructure at the beginning of deformation at time $t = 0$.
Integration by parts results in
\begin{equation}
    E_\mathrm{stored}(t) = E_\infty(t) - \bigg\{ E_\infty(0) - E_{0} + \int_0^t \deriv{E_\infty}{t'} \exp[\zeta W_{p}(t')] dt' \bigg\} \exp[-\zeta W_{p}(t)].
    \label{eq:E_stored_model}
\end{equation}
Let $I(t)$ represent the integral in Eq.\ \ref{eq:E_stored_model}, and observe that $I(t)$ vanishes for fixed straining conditions.
Substituting into Eq.\ \ref{eq:beta_diff} gives the following explicit form for the differential TQC:
\begin{align}
    \beta_\mathrm{diff}(t) &= 1 - d E_\mathrm{stored}(t) / d W_p(t) \nonumber \\
    &= 1 - \zeta [ E_\infty(0) - E_{0} + I(t) ] \exp[-\zeta W_{p}(t)].
    \label{eq:beta_diff_model}
\end{align}
Similarly, substituting into Eq.\ \ref{eq:beta_int} gives the following explicit form for the integral TQC:
\begin{align}
    \beta_\mathrm{int}(t) &= 1 - [E_\mathrm{stored}(t) - E_0] / W_p(t) \nonumber \\
    &= 1 - \{ E_\infty(t) - E_0 - [ E_\infty(0) - E_{0} + I(t) ] \exp[-\zeta W_{p}(t)] \} / W_p(t).
    \label{eq:beta_int_model}
\end{align}

To fully define the solution, we assume that the asymptotic value of the stored energy has the form $E_\mathrm{\infty}(t) = a \dot{\epsilon}_p^b(t)$ where $\dot{\epsilon}_{p}$ is the plastic strain rate.
This is consistent with the MD simulations performed here where the strain rate is the only control variable that substantially changes in time, with temperature, pressure and crystal orientation all remaining effectively constant.
This leads to 
\begin{equation}
I(t) = a b \int_0^t \dot{\epsilon}_p^{b - 1}(t') \deriv{\dot{\epsilon}_p}{t'} \exp[\zeta W_{p}(t')] dt'
\end{equation}
as the explicit form of the integral in Eqs.\ \ref{eq:E_stored_model}, \ref{eq:beta_diff_model} and \ref{eq:beta_int_model}.
As mentioned above, the constant plastic strain rate in simulation A and the second part of simulation B (after the rate ramping) causes the contribution to the integral $I(t)$ to vanish and makes the solution particularly simple during those intervals.
The solution for the first part of simulation B from $0\ \mathrm{ns}$ to $2\ \mathrm{ns}$ during which the strain rate is linearly increasing is only slightly more complicated, and effectively entails numerically integrating the time derivative of the plastic strain rate and the exponential of the plastic work.
The solution for the stored energy value attained at the end of the rate ramping interval serves as the initial value $E_0$ for the subsequent part of simulation B during which the strain rate is held constant at $4 \times 10^8\ \mathrm{s^{-1}}$.

\begin{figure}
	\centering
	\includegraphics[width=0.45\textwidth]{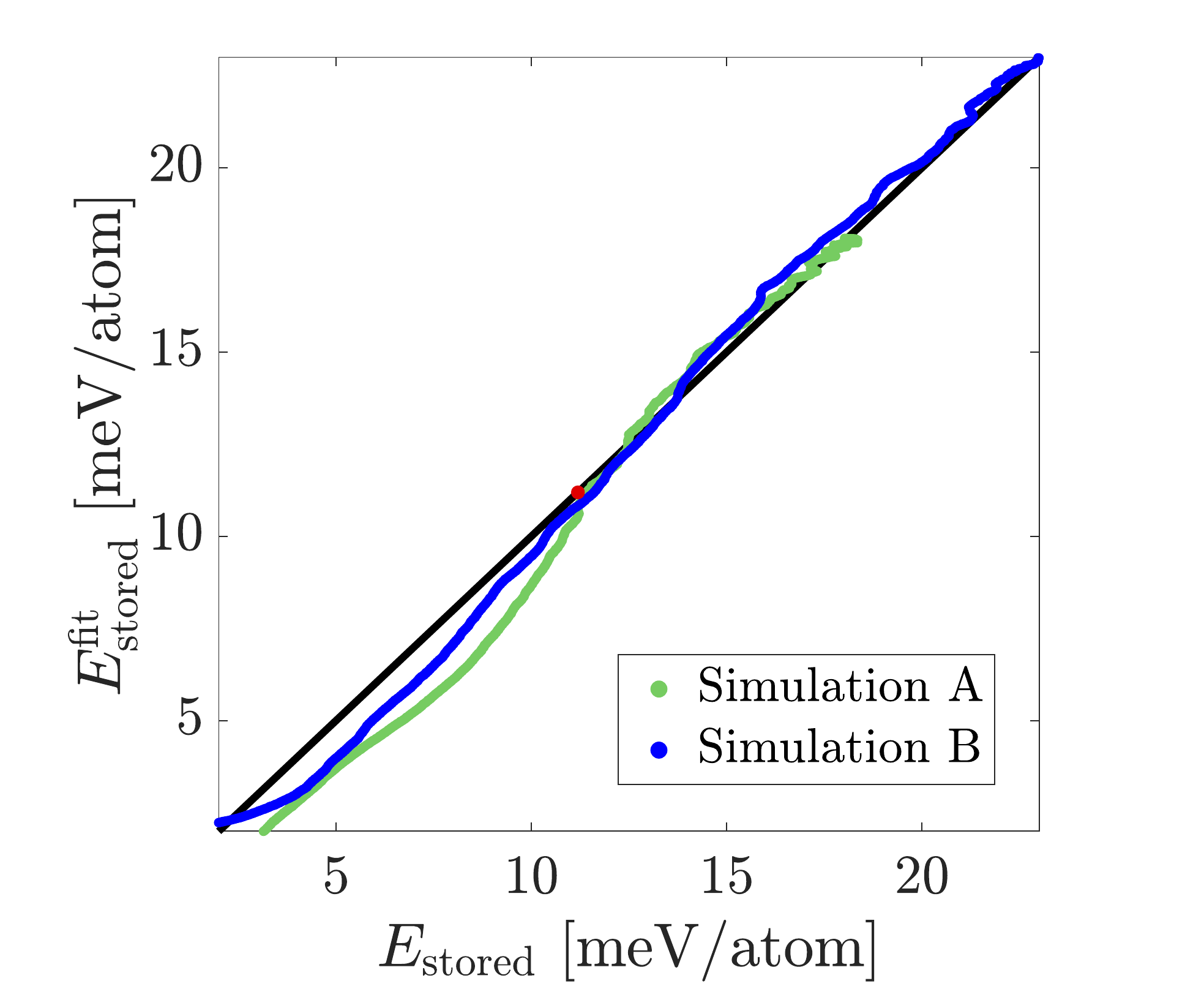}
	\caption{Comparison of the predictions of Eq.\ \ref{eq:E_stored_model}, $E_\mathrm{stored}^\mathrm{fit}$, with the energy stored in the defect microstructure, $E_\mathrm{stored}$, for both simulations.
	The red marker approximately indicates the point where steady-state plastic flow begins in simulations A and B.
	The overall RMSE is $0.502\ \mathrm{meV / atom}$.}
	\label{fig:phenom_energy}
\end{figure}

\begin{figure}
	\includegraphics[width=0.90\textwidth]{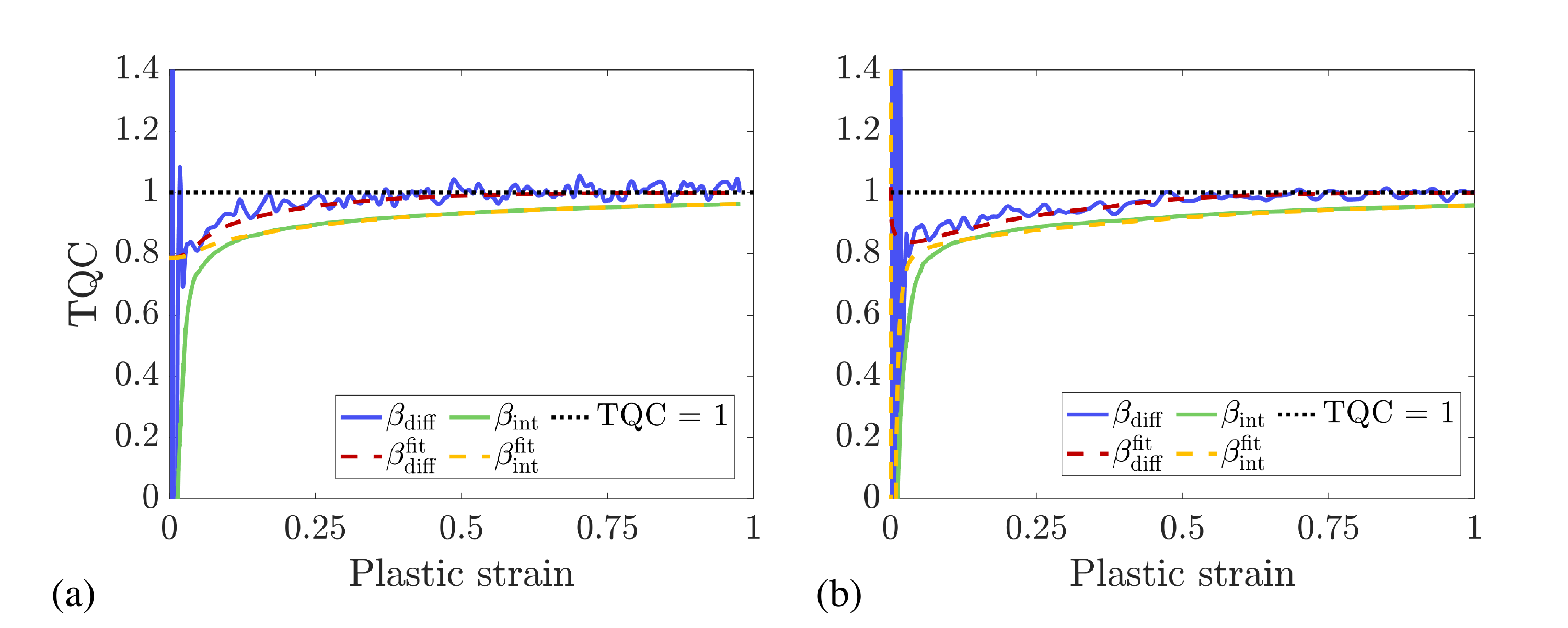}
	\caption{Comparison of the predictions of Eqs.\ \ref{eq:beta_diff_model} and \ref{eq:beta_int_model} with the values of $\beta_{\mathrm{diff}}$ and $\beta_{\mathrm{int}}$ plotted in Fig.\ \ref{fig:TQC} for (a) simulation A and (b) simulation B.}
	\label{fig:phenom_TQC_diff}
\end{figure}

Let $E_A$ and $E_B$ be the stored energy values asymptotically reached in the $t \rightarrow \infty$ limit while holding the strain rate constant at $\dot{\epsilon}_A = 2 \times 10^8\ \mathrm{s^{-1}}$ and $\dot{\epsilon}_B = 4 \times 10^8\ \mathrm{s^{-1}}$, respectively.
$a$, $b$, $\zeta$, $E_A$ and $E_B$ could be used as five parameters to fit the phenomenological model to the MD simulation data for stored energy and plastic work, though the assumed form for $E_\mathrm{\infty}(t)$ reduces the number of independent parameters to three with
\begin{equation}
    a = \frac{E_B - E_A}{\dot{\epsilon}_B^b - \dot{\epsilon}_A^b} \qquad
    b = \frac{\log E_B - \log E_A}{\log \dot{\epsilon}_B - \log \dot{\epsilon}_A}.
\end{equation}
We used a standard nonlinear least squares procedure to obtain the values for the three independent parameters $\zeta$, $E_A$ and $E_B$ that provide the best fit to the combined $E_\mathrm{stored}$ data for simulations A and B, with the resulting values being $\zeta = 0.0126 \ \mathrm{atom / meV}$, $E_A = 18.1\ \mathrm{meV / atom}$, and $E_B =23.4\ \mathrm{meV / atom}$.
The fit quality is illustrated in Fig.\ \ref{fig:phenom_energy} where the $E_\mathrm{stored}$ data are plotted in different colors for simulations A and B against the predictions of the phenomenological model.
The accuracy of the model noticeably improves for both simulations after steady-state plastic flow is reached (indicated by the red marker in the figure), likely as a consequence of the model not being able to capture the complex processes of defect multiplication around the yield point.
The same parameters were used to evaluate the predictions of Eqs.\ \ref{eq:beta_diff_model} and \ref{eq:beta_int_model} for $\beta_{\mathrm{diff}}$ and $\beta_{\mathrm{int}}$, with the results shown in Fig.\ \ref{fig:phenom_TQC_diff}.
The model agrees reasonably well with the values of TQC extracted directly from the two MD simulations, with the most significant deviations occurring right after yielding for plastic strains of less than $0.1$;
this is the same interval where the fluctuations in the observed values of the TQC are largest.

Fundamentally, the TQC reflects the complexity of microstructural evolution \cite{2019Nieto-Fuentes} and should depend not only on the strain rate \cite{2007Rittel} but additionally on temperature, pressure, crystal orientation \cite{2009Rittel}, and even loading mode \cite{2017Rittel}.
A phenomenological model of the kind described here, while not resolving any relevant details of the microstructure evolution, offers an approximate quantitative description that goes beyond the current practice where TQC is treated as a material constant.
Indeed, at the expense of three adjustable parameters, such a model is likely to predict with better accuracy the rate of heat release and the resulting rise in temperature as a function of time.
If the parameter $\zeta$ is indeed a material constant, that would allow the kinetics of energy storage to be adequately described over complex straining trajectories with variations in pressure, temperature and loading mode provided that the functional form of the asymptotic stored energy $E_{\mathrm{\infty}}(t)$ is known.
Given the dependence of material response on released heat and temperature, this would considerably reduce experimental uncertainties about material behavior under extreme straining conditions.

\subsection{Stored energy model}
\label{subsec:stored_energy}

Having defined stored energy as a measurable macroscopic quantity, the purpose of this section is to identify microscopic defects that actually contribute to $E_\mathrm{stored}$.
Here we show that, although the majority of the stored energy is attributed to dislocations, the dislocation contribution alone does not account for all of the energy storage in our MD simulations.
It is significant that an early review on the TQC included the contributions of point defects, along with dislocations, when estimating the stored energy of cold work \cite{1973Bever}.
Indeed, our MD simulations indicate that point defects including vacancies, interstitials, and their clusters contribute substantially to the energy stored in the defect microstructure.
The concentration of point defects generally increases with increasing strain rates and dislocation density. The defects are produced largely as debris in the wake of moving screw dislocations dragging cross-kinks and/or jogs formed by dislocation intersections \cite{anderson2017theory} \cite{marian2004dynamic}.
Moreover, in the absence of internal interfaces or other sinks, point defect recombination and absorption by dislocations are the two most likely mechanisms of point defect removal in the crystal bulk.

The copious point defect debris in our high strain rate simulations is mostly single vacancies, though single interstitials and interstitial clusters also occur.
The BCC Defect Analysis (BDA) method \cite{Moller2016BDA} was used to count the vacancies produced in our simulations.
The same BDA method is inaccurate in counting interstitials, even though we were able to visually identify multiple interstitials and interstitial clusters using various functionalities of the OVITO software \cite{ovito}.
Visually, interstitials appear to be less numerous than vacancies, but their contribution to energy storage should not be discounted because the formation energy of a single interstitial in the Ta model employed here is $7.54 \ \mathrm{eV}$, nearly three times that of a vacancy at $2.75 \ \mathrm{eV}$.
We assume that energy stored in the interstitials is directly proportional to the energy stored in the vacancies at all times.

\begin{figure}
	\centering
	{\includegraphics[width=0.90\textwidth]{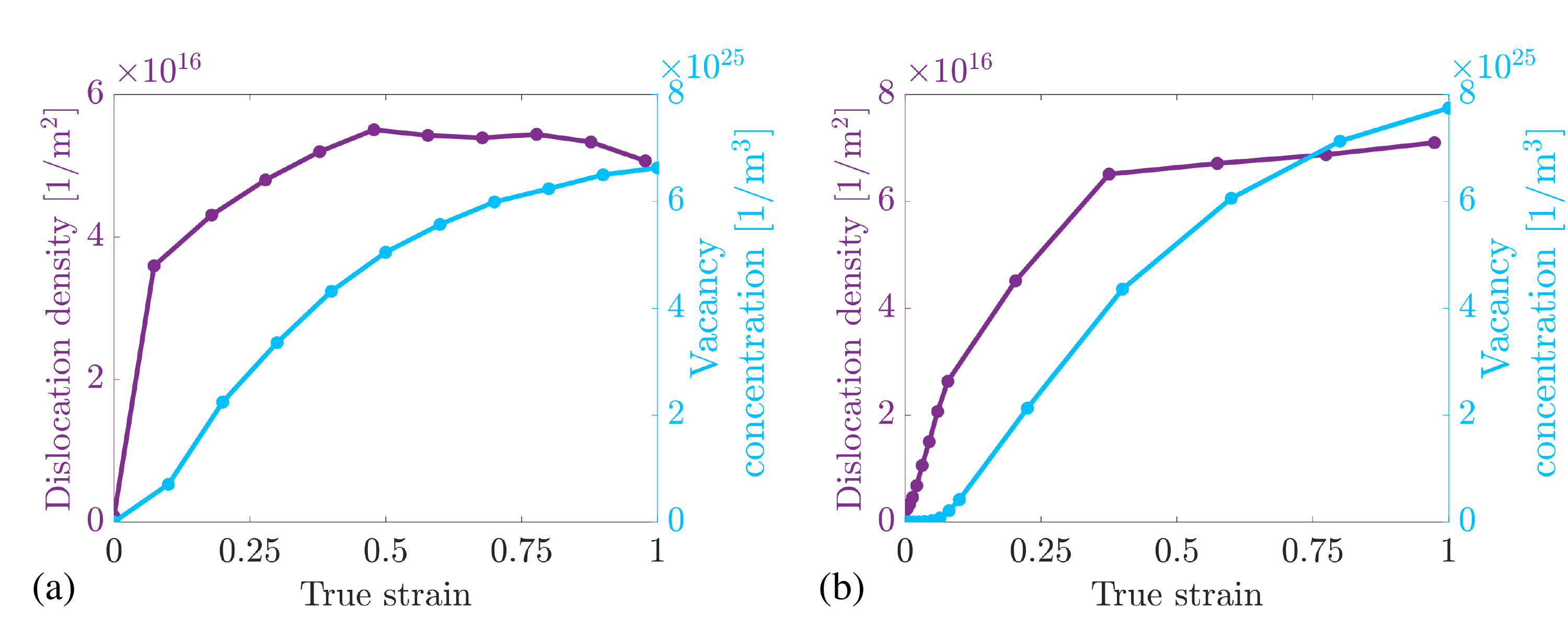}}
	\caption{Dislocation density and vacancy concentrations as functions of true strain for (a) simulation A and (b) simulation B.}
	\label{fig:Dislocation_and_vcy_dens}
\end{figure}

Shown in Fig.\ \ref{fig:Dislocation_and_vcy_dens}a is a plot of the vacancy concentration in simulation A as a function of strain along with the evolution of the dislocation density extracted using DXA \cite{Stukowski10, Stukowski14}.
At a strain of $\about 0.5$ where the dislocation density appears to have saturated, the vacancy concentration is still rising and has not fully saturated even by the end of simulation A at a strain of $1.0$.
The same trend can be seen in Fig.\ \ref{fig:Dislocation_and_vcy_dens}b pertaining to simulation B.
This asynchronicity could reflect the fact that, while the rate of debris production depends only on dislocation density, the rate of debris removal should also depend on the point defect concentration itself. 
Thus, for the rate of debris removal to balance the rate of debris production by dislocations, the debris may need to reach sufficiently high concentrations that are not fully attained even at a strain of $1.0$.
Assuming that at any given time the energy of stored interstitials is proportional to the energy of stored vacancies, the contribution of the debris to the stored energy is modeled by multiplying the energy stored in vacancies by a constant factor $\xi > 1$:
\begin{equation}
E_{\mathrm{debris}} = \xi N_{\mathrm{vcy}} E_{\mathrm{vcy}}
\label{eq:Edebris}
\end{equation}
where $E_{\mathrm{vcy}}$ is the vacancy formation energy and $N_{\mathrm{vcy}}$ is the number of vacancies counted using the BDA algorithm. 

The energy stored in the dislocation network is modeled following Ref.\ \cite{bertin2018}:
\begin{equation}
E_{\mathrm{dis}} = \eta \chi \frac{ \mu b^2}{4 \pi} \rho \ln \left(\frac{1}{r_c \sqrt{\rho}}\right)
\label{eq:E_dis}
\end{equation}
which includes the linear dependence of the elastic energy on the dislocation density $\rho$ and a logarithmic term that accounts for the screening and core cutoff $r_c$ of the dislocations' stress fields.
Here $\mu$ is the shear modulus, $b = 0.286\ \mathrm{nm}$ is the Burgers vector magnitude, and $\chi$ and $\eta$ are parameters on the order of unity that are discussed below and in more detail in the  Appendix.
Although this model was initially introduced to describe the energy associated with arrays of infinite straight dislocations, it has been recently shown to accurately capture the energetics of complex and realistic dislocation networks \cite{bertin2018}.

To more accurately account for the energy stored in dislocations, Eq.~\ref{eq:E_dis} is adjusted from Ref.\ \cite{bertin2018} as follows. 
First, a factor $\eta$ is introduced to correct the isotropic model proposed in Ref.\ \cite{bertin2018} for the elastic anisotropy of the Ta model used in our MD simulations.
As described in the Appendix, this is estimated to be $\eta \simeq 1.35$ at $300\ \mathrm{K}$.
Second, a factor $\chi$ is used as a fitting parameter to account for variations in the stored energy as a consequence of the precise geometric arrangement of the dislocation lines.
Although the physical meaning of $\chi$ is subtle, as was shown in Ref.\ \cite{bertin2018} and is further discussed in the Appendix, physically reasonable values of this parameter are bound to the interval between $1.0$ and $1/(1-\nu) \simeq 1.5$.
The cutoff radius is set to $r_c = 0.5 b$ as a coarse estimate for dislocation core energies reported for the same model of Ta elsewhere \cite{bertin2021core}.
More details on the parameters $\eta$, $\chi$, $r_c$, and their relationships are given in the Appendix.

\begin{figure}
	\centering
    \includegraphics[width=0.9\textwidth]{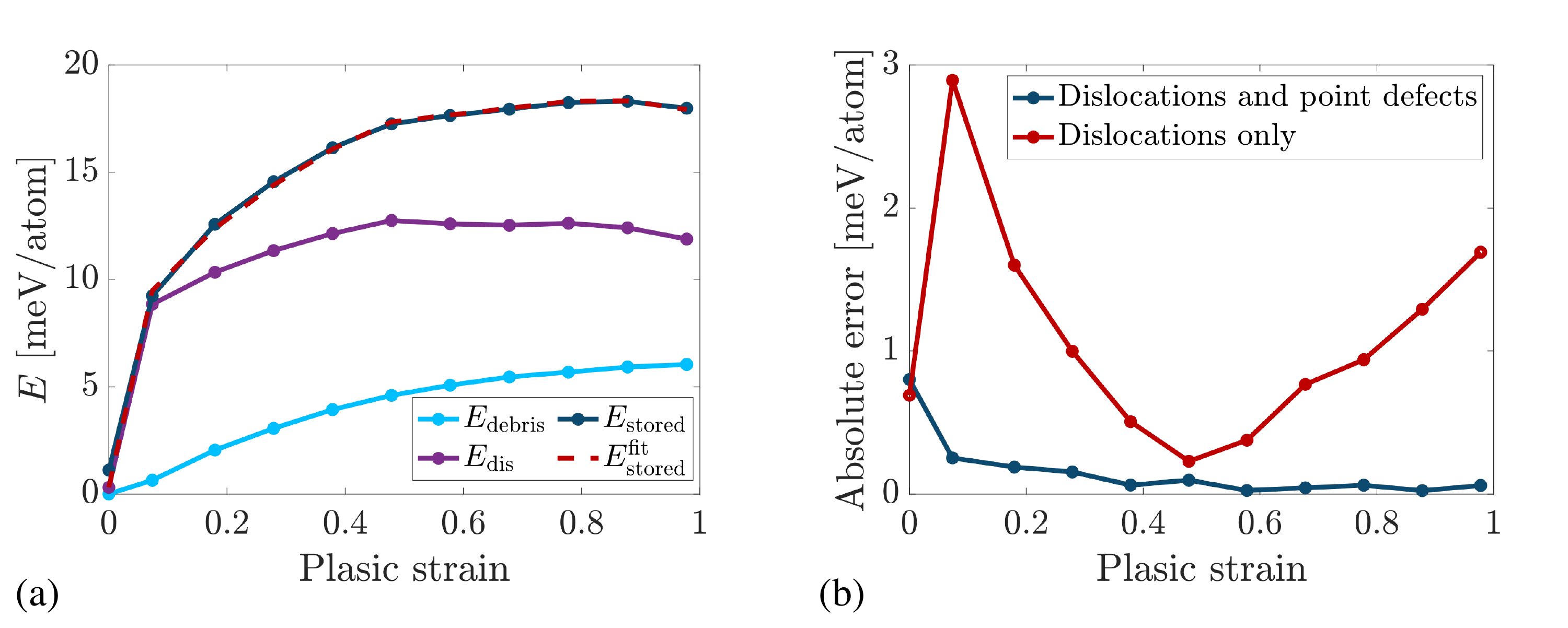}
	\caption{Stored energy model for simulation A as a function of true plastic strain.
	(a) The model that includes dislocations and point defects gives $\chi = 1.26$ and $\xi = 1.73$, with the model and the separate contributions of the two defect terms shown.
	(b) The absolute error for the model in (a), as well as for a version that excluded the point defect contribution which resulted in $\chi = 1.72$.}
	\label{fig:energy_model_A}
\end{figure}

\begin{figure}
	\centering
	\includegraphics[width=0.90\textwidth]{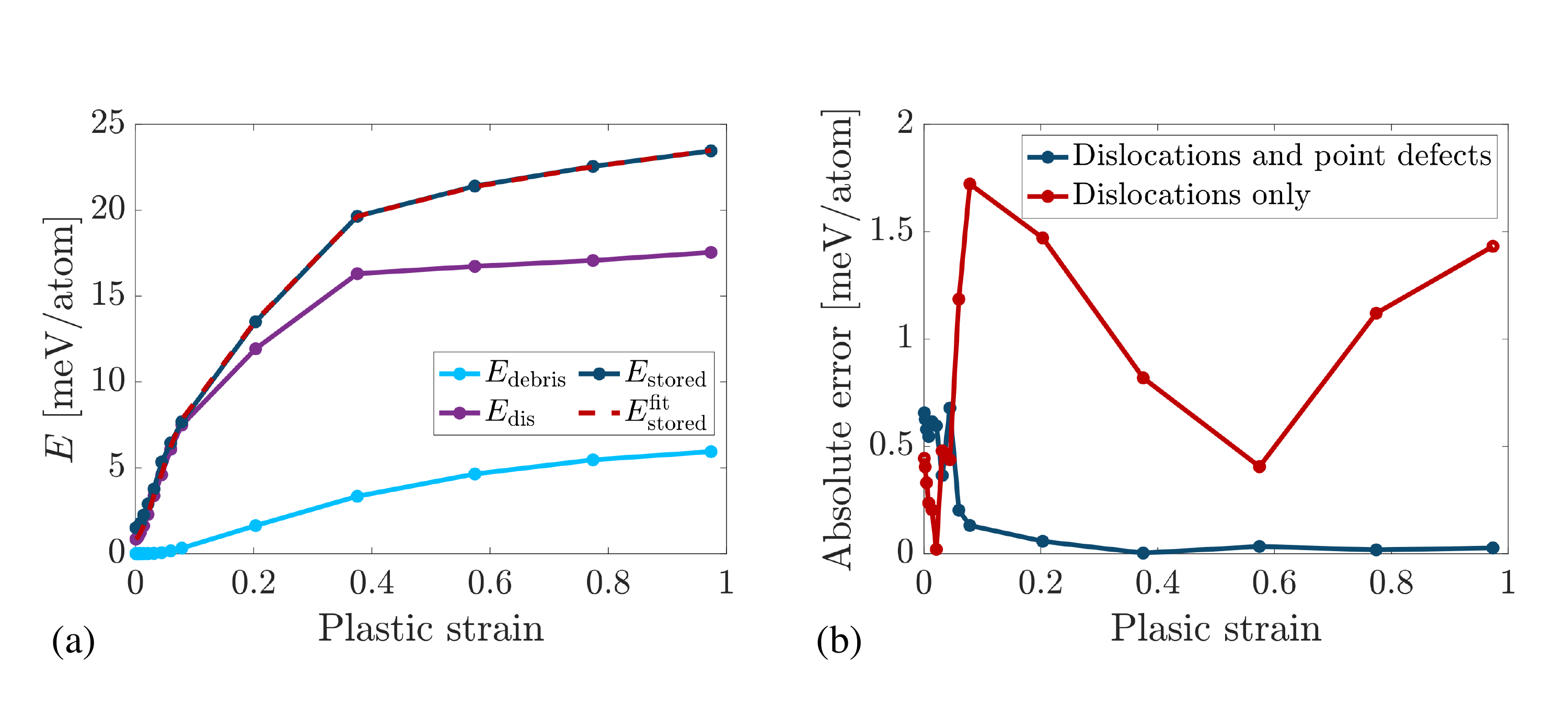}
	\caption{Stored energy model for simulation B as a function of true plastic strain.
	(a) The model that includes dislocations and point defects gives $\chi = 1.40$ and $\xi = 1.55$, with the model and the separate contributions of the two defect terms shown.
	(b) The absolute error for the model in (a), as well as for a version that excluded the point defect contribution which resulted in $\chi = 1.75$.}
	\label{fig:energy_model_B}
\end{figure}

Assuming that the point defect debris and the dislocations are the only contributions to the energy of the defect microstructure gives
\begin{equation}
E_{\mathrm{stored}} = E_{\mathrm{dis}} + E_{\mathrm{debris}}
\label{eq:E_fit}
\end{equation}
where the coefficients $\xi$ and $\chi$ in Eqns.~\ref{eq:Edebris} and \ref{eq:E_dis}, respectively, are treated as fitting parameters to describe the variations in the stored energy data  $E_{\mathrm{stored}}$ extracted from our MD simulations.
Results for the fit obtained with Eq.~\ref{eq:E_fit} are shown in Fig.~\ref{fig:energy_model_A}a for simulation A, with $\chi = 1.26$ and $\xi = 1.73$ providing a remarkably good fit to the energy storage data.
The value of $\xi > 1$ indicates that the contribution to the stored energy from the unknown population of interstitials is a significant fraction of that of the vacancies, while the value of $\chi = 1.26$ falls within the expected interval defined elsewhere \cite{bertin2018}.
Shown in Fig.~\ref{fig:energy_model_A}b is the resulting absolute error for the fit in Fig.~\ref{fig:energy_model_A}a and the error for the best fit with $\xi = 0$, ignoring the contribution of the point defects and accounting only for the dislocation network energy.
Not including point defects results both in a best fit value $\chi = 1.72$ which is outside the acceptable range for this parameter
and significantly higher absolute error, supporting our assessment that the dislocation network alone cannot reasonably account for the total stored energy.
Thus, we find that point defects provide a significant mechanism for energy storage, with their contribution potentially half that of the (very dense) dislocation network at strains of $1.0$ in Fig.~\ref{fig:energy_model_A}a.
Similar results were obtained for simulation B in Fig.~\ref{fig:energy_model_B}a for which the best fit values are $\chi = 1.40$ and $\xi = 1.55$.
Excluding the contribution of point defects gives $\chi = 1.75$ which, just as for simulation A, falls outside the physically admissible range.
The absolute errors of the model, both for the version that includes the contribution of point defects and for the one that does not, are provided in Fig.~\ref{fig:energy_model_B}b. 
That the best fit parameters have slightly different values for simulations A and B likely reflects differences in their straining trajectories and ultimate strain rates.

\section{Conclusion}
\label{sec:conclusion}

Taylor and Quinney defined a thermo-mechanical conversion factor (TQC) as a measure of the energy stored in a material undergoing plastic deformation.
This continues to be a quantity of enduring interest that can be and has been experimentally measured at low and moderately high deformation rates.
That said, fully dynamic high strain rate simulations of the kind reported here can be used to quantify thermomechanical conversion under extreme deformation conditions where experimental measurements are difficult or impossible.  

Since the original study of Taylor and Quinney, advances in microscopy and material characterization have offered a variety of ways to observe and quantify changes in the defect microstructure that contribute to energy storage during mechanical deformation.
Likewise, the atomistic simulations reported here allow one to observe the mechanisms of energy storage during high-rate deformation \emph{in silico} and in arbitrary detail.  

Our results support the following conclusions:
\begin{itemize}
    \item Under compressive deformation conditions held unchanged for sufficient strains, single crystal tantalum is predicted to asymptotically approach a state of steady flow in which the defect microstructure becomes statistically stationary.
    Energy storage ceases and the TQC approaches $1.0$ in the same asymptotic limit.
    \item The asymptotic limit $E_\infty$ of stored energy is a measure of the material's energy storage capacity and can be computed and tabulated as a function of the straining conditions.
    It is used here to define a phenomenological model of energy storage kinetics potentially applicable to complex deformation histories.   
    \item Dislocation multiplication is a major energy storage mechanism in the high-rate deformation conditions studied here.
    However, point defect debris is also produced in copious quantities and contributes as much as half of the energy stored in the dislocation network.
    Dislocation multiplication and debris production together account for the TQC staying measurably below $1.0$ up to large compressive strains.
    \item Although minor, energy storage due to debris production continues well beyond the strains at which dislocation multiplication ceases to contribute to energy storage. 
\end{itemize}

\begin{acknowledgments}

The authors acknowledge useful discussions with D. Swift, S.\ Aubry, N.\ Barton and D.\ Mordehai.
J.C.S.\ was partially supported by the Hellman Family Foundation.
This work was performed under the auspices of the U.S.\ Department of Energy by Lawrence Livermore National Laboratory under contract DE-AC52-07NA27344.

\end{acknowledgments}

\section*{Author contributions}
\label{sec:contributions}

JCS and NB performed all simulations.
All authors developed the models and concepts, analyzed simulation results and wrote the paper.

\section*{Competing interests}
\label{sec:interests}

The authors declare no competing interests.

\bibliography{references}

\appendix

\section{Dislocation network energy model}
\label{sec:appendix}


This section provides additional details on the model used for the evaluation of the dislocation network energy in our Ta simulations. Following Ref.\ \cite{bertin2018}, the energy of a complex dislocation network in an isotropic medium is well described by the expression
\begin{equation}
E_{\mathrm{dis}} = C \frac{\mu b^2}{4 \pi} \rho \ln \left(\frac{1}{r_c \sqrt{\rho}}\right)
\label{eq:E_dis_app}
\end{equation}
where $\mu$ is the shear modulus, $b$ is the Burgers vector magnitude, $\rho$ is the dislocation density, and $r_c$ is the core cutoff radius.
$C$ is a coefficient that depends on the distribution of the dislocation line character angles, with values ranging from $C = 1.0$ for an array of pure screw dislocations to $C = 1/(1-\nu)$ for an array of edge dislocations, where $\nu$ is Poisson's ratio.
Reference \cite{bertin2018} examined the validity of Eq.~\ref{eq:E_dis_app} for complex dislocation networks using discrete dislocation dynamics (DDD) simulations in FCC crystals and showed that this simple model is accurate with the value of $C$ reflecting the dislocation character angle averaged over all lines in the network.

\begin{figure}
	\centering
	{\includegraphics[width=0.90\textwidth]{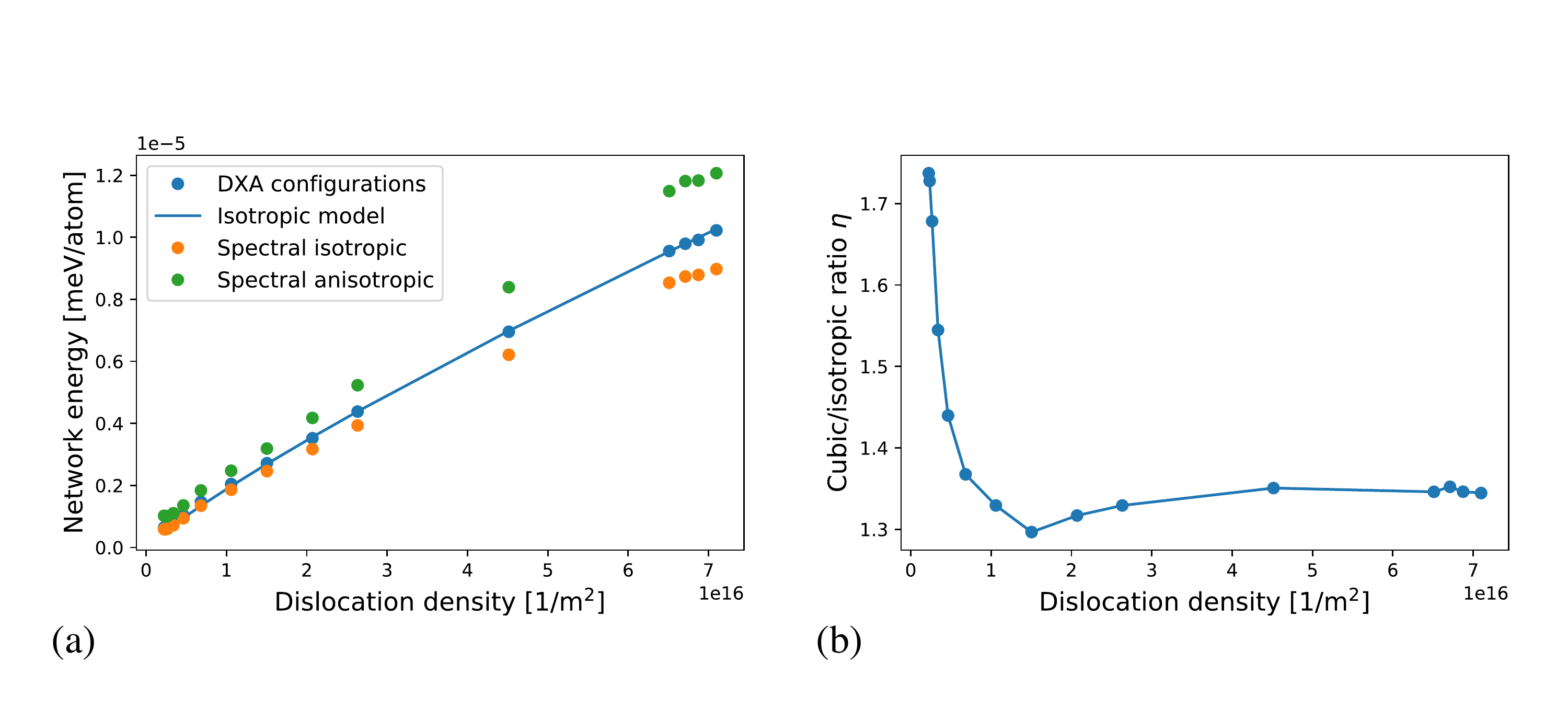}}
	\caption{(a) Evaluation of the dislocation network energy as a function of the dislocation density using different methods and the isotropic model in Eq.~\ref{eq:E_dis_app}.
	(b) Evolution of the empirical anisotropic correction factor $\eta$ defined as the ratio of the energies computed for cubic and isotropic elasticity using the spectral method.}
	\label{fig:dxa_energy}
\end{figure}

Here the validity of the same analytical model is examined for dislocation networks in Ta.
For this purpose we first extracted the dislocation networks attained at several strain values in simulation B using the DXA method \cite{Stukowski10, Stukowski14}.
Energies of the extracted dislocation networks were found by summing up the pairwise interaction energies of all dislocation segments in the networks as in Ref.\ \cite{bertin2018}. 
Since real space solutions for such interaction energies are only available for elastically isotropic crystals, isotropic elasticity was assumed with $\mu = 55$ GPa and $\nu= 0.339$. 
The core energy contributions were computed for the same networks using a simple core model described in Ref.\ \cite{bertin2018} and added to the elastic energies. The so-computed network energies containing both isotropic elastic and core contributions are plotted in Fig.~\ref{fig:dxa_energy}a as a function of the dislocation density.
On the same plot the best fit to the analytical energy model in Eq.~\ref{eq:E_dis_app} is also shown as a solid blue line with $C$ as the only fitting parameter.
This simple model captures the energy of our complex BCC networks quite well.
The best fit was obtained for $C = 1.1$ which is consistent with the average character angle of $\theta \simeq 20^o$ computed from the fully developed dislocation networks in our MD simulations;
this average angle reflects the prevalence of screw dislocations in our configurations.


Despite seemingly working well, the simple analytical model in Eq.~\ref{eq:E_dis_app} needs to be corrected to be applicable to the dislocation networks produced by our MD simulations.
First, variations in the core energy with dislocation character angle are subtle and, as was shown in Ref.\ \cite{bertin2021core}, cannot possibly be described using a single value of the core cutoff parameter $r_c$.
However, this inaccuracy only enters through the slowly-varying logarithmic term.
Fixing the core radius at $r_c = 0.5b$ is generally consistent with the core energy values calculated at $0\ \mathrm{K}$ in Ref.\ \cite{bertin2021core}, and all remaining variations are subsumed into the single fitting parameter $\chi$.
For simplicity, the same parameter $\chi$ also accounts for the character angle dependence of the network energy.
We expect $\chi$ to be in the range $1 \leq \chi \leq 1/(1-\nu)$, similar to the previously-used fitting parameter $C$. 

A second and more substantial correction is required to account for the elastic anisotropy of our atomistic Ta model which has a Zener's anisotropy ratio of $2 C_{44} / (C_{11} - C_{12}) = 2.1$ at $300\ \mathrm{K}$.
This could make the approximation of isotropic elasticity previously used to validate Eq.~\ref{eq:E_dis_app}, and in particular the $\mu b^2 / (4 \pi)$ energy factor, significantly inaccurate.
This inaccuracy is accounted for using an anisotropy correction factor $\eta$.
Computing $\eta$ would involve computing the energies of the same dislocation networks in a crystal with cubic symmetry.
Since there are no analytical solutions for the interaction energies of dislocation segments in cubic crystals, we resort to a spectral method in which the energy of the elastic field induced by a dislocation network is computed by directly integrating the strain energy density over the simulation volume \cite{bertin2019connecting}.

The method entails summing up Fourier components of the elastic strain energy obtained using the elastic Green's function available in an analytical form for solids of arbitrary symmetry in the $k$-space.
To obtain the correction factor, this spectral method is used to compute and compare energies of the same dislocation networks with full cubic anisotropy and in the isotropic approximation.
The spectral method is approximate, with an accuracy limited by the spacing of the grid used by the Fast Fourier Transform (FFT) to map the field values into $k$-space.
The spectral method was used to compute both the cubic and the isotropic energies for a number of dislocation networks extracted from simulation B for grids containing $128^3$, $256^3$ and $512^3$ points.
These grids are likely too coarse to fully capture the variations of the strain fields near the dislocation cores, as indicated by the results for the $256^3$ grid shown in Fig.~\ref{fig:dxa_energy}a (orange and green symbols).
Compared to the energies computed using the more accurate real-space method of Ref.\ \cite{bertin2018} (blue symbols), the network energies computed with the spectral method in the isotropic approximation are consistently lower, a consequence of the smearing of the elastic strain field on the regular grid.
Nevertheless, we observe that the ratio of the cubic to the isotropic network energies is nearly independent of the grid size.
At higher dislocation densities where the networks are fully developed, the ratio of the anisotropic and isotropic energies is found to converge to a value $\eta \simeq 1.35$ (Fig. \ref{fig:dxa_energy}b).
This is introduced as the correction factor in 
\begin{equation}
E_{\mathrm{dis}} = \eta \chi \frac{\mu b^2}{4 \pi} \rho \ln \left(\frac{1}{r_c \sqrt{\rho}}\right),
\end{equation}
and is subsequently used to fit the above equation to the energy storage data extracted directly from our MD simulations.

\end{document}